\documentclass[a4paper, fleqn]{article}

 \usepackage{bm}
 \usepackage{amssymb}
 \usepackage{latexsym}
 \usepackage{amsfonts}
 \usepackage{epsfig}
 \usepackage{color}
 \definecolor{darkblue}{rgb}{0,0,.5}
 \usepackage[linktocpage, colorlinks=true ,linkcolor=darkblue, citecolor=darkblue]{hyperref}
 \usepackage[all]{hypcap}

 \newcommand{\expval}[1]{\left< #1 \right>}
 \newcommand{\dexpval}[1]{\left<\left< #1 \right>\right>}
 
\newcommand{\ket}[1]{\left|#1\right>}
\newcommand{\bra}[1]{\left<#1\right|}
 
 \newcommand{\nn}{\nonumber\\}
 
 \newcommand{\f}[1]{\mbox{\boldmath$#1$}}

 \newcommand{\bea}{\begin{eqnarray}}
 \newcommand{\ea}{\end{eqnarray}}
 \newcommand{\eea}{\end{eqnarray}}
 \newcommand{\ord}{{\cal O}}
 \newcommand{\trace}[1]{{\rm Tr}\left\{ #1 \right\}}

 \newcommand{\traceS}[1]{{\rm Tr}\left\{ #1 \right\}}
 \newcommand{\abs}[1]{{\left| #1 \right|}}

 \newcommand{\ii}{{\rm i}}

\begin{document}


\title{Counting statistics of collective photon transmissions}

 \author{M. Vogl$^*$,\\
\and G. Schaller$^\dagger$,\\
\and T. Brandes\\
\small $^*$ malte.vogl@tu-berlin.de,\\
\small $^\dagger$ gernot.schaller@tu-berlin.de,\\
Institut f\"ur Theoretische Physik, Technische Universit\"at Berlin,\\
Hardenbergstr. 36, 10623 Berlin, Germany
}

\maketitle
\begin{abstract}
  We theoretically study cooperative effects in the steady-state transmission 
  of photons through a medium of $N$ radiators.
  Using  methods from quantum transport, we find 
  a cross-over in scaling from $N$ to $N^2$ in the current and to even higher powers of $N$ in the higher cumulants of the 
  photon counting statistics as a function of the tunable source occupation. 
  The effect should be observable for atoms confined within a nano-cell 
  with a pumped optical cavity as photon source. 
\end{abstract}

\section{Introduction} 
Collective effects in the emission of light 
have attracted a lot of attention recently. Super-radiant scattering off Bose-Einstein condensates~\cite{ICSSPK99},
collective self-organization of atoms~\cite{DR02,BCV03}, or the observation of the phase transition~\cite{HL73}
for ultra-strong coupling in a cavity~\cite{BGBE10} are some of the modern manifestations of the original 
Dicke effect~\cite{Dic53,Dic54}.

There has also been an increased interest in identifying 
similar collective interference phenomena in mesoscopic (electronic) transport~\cite{Bra05}.
Many of these activities are triggered by progress in the time-resolved detection and counting of individual electrons~\cite{FCS_exp}, and 
the fabrication of quantum circuits as electronic test-beds for quantum optical effects~\cite{cQEDreview}.

With the progress obtained in single-photon detectors~\cite{hadfield2009a} we feel that it is well motivated to 
revert the focus by applying theoretical methods from Full Counting Statistics (FCS)~\cite{Naz03}
to steady-state transmission of photons through a medium of $N$ radiators in the small-sample limit of the original super-radiance model~\cite{GH82}. 
We show that higher cumulants of the photon counting statistics are sensitive indicators of 
cooperative emission effects even for only moderate pumping of the radiators.

Steady-state super-radiance of incoherently pumped atoms in a cavity has been analyzed recently by Meiser and co-workers in terms of 
time-dependent intensity correlation functions~\cite{MH10}. 
Our results provide an alternative scheme, where we propose to confine atoms within a nano-cell
and to use a pumped optical cavity as photon source.
The radiation emitted by the atoms is analyzed in terms of cumulants of a stationary photo-detection statistics.
Instead of directly evaluating the cumulant-generating function -- this is analytically impossible for the large 
system sizes considered -- we extract the long-term cumulant behavior from the Laplace transform of the 
moment-generating function.
We also expect our findings to be relevant for 
other situations  dealing with collective boson transport, such as thermal transport 
\cite{Cahillaetal03} or phonon lasers~\cite{grudinin2010}. 
%
 %
 \section{Model} 
\begin{figure}[b]
 \begin{center}
 \includegraphics[width=0.4\textwidth,clip=true]{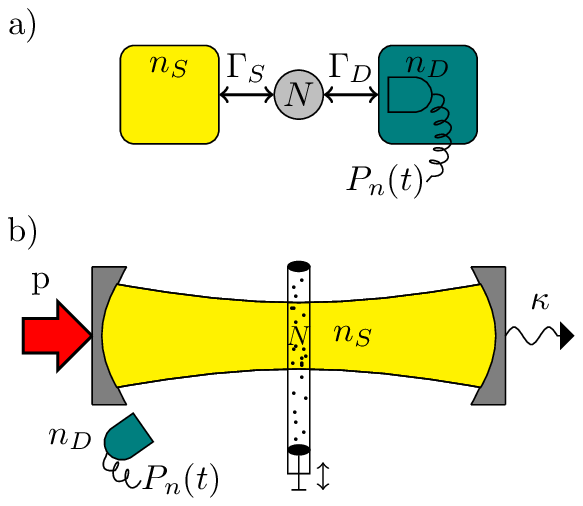}
 \end{center}
 \caption{(Color Online)
 a) Excitation transfers between a medium of $N$ non-interacting two-level systems  and $N_\alpha=2$ bosonic reservoirs
 (source $S$ and drain $D$ with occupation $n_S>n_D$) occur at rates $\Gamma_{S/D}$
 and are tracked by counting devices.
 b) Laser-driven optical multi-mode cavity with pump (loss) rate $p (\kappa)$ containing 
 $N$ atoms that collectively radiate into non-cavity modes.
 Nanometer-sized gas cells constrain the atoms in two dimensions and allow for a variation of the atom number $N$ in the 
 interaction region.
 }
 \label{fig:sketch}
 \end{figure}

 We consider a number of $N_\alpha$ bosonic reservoirs $\alpha$ connected via  
 a medium of $N$ two-level systems with identical level splittings $\Omega$. The collective coupling 
 is described by an extended multi-mode Dicke Hamiltonian (cf. Figure~\ref{fig:sketch} a),
 \bea\label{Ehamiltonian}
 H &=&\frac{\Omega}{2} J^z +\sum_{k,\alpha} \left[\omega_{k\alpha} b_{k\alpha}^{\dagger} b_{k\alpha} + \left(J^x h_{k\alpha} b_{k\alpha}^{\dagger} +{\rm h.c.}\right)\right]
 \eea
 with collective  
 pseudo-spin operators
  $J^{x,z}\equiv\sum_{i=1}^N \sigma^{x,z}_i$. Here,  $b^{\dagger}_{k\alpha}$ creates 
 a boson mode $k$ with frequency $\omega_{k\alpha}$ in reservoir $\alpha$
 and $h_{k\alpha}$ denote the coupling constants.
We derive the counting statistics of the total number $n$  of bosons exchanged between the medium and one ($\bar\alpha$) of the reservoirs 
by formally introducing~\cite{SKB09}
a counting operator \mbox{$d_{\bar\alpha}^{\dagger}= \sum_{n=-\infty}^{+\infty} \ket{n+1}\bra{n}$} (and similarly for the reverse tunneling processes)
in the Hamiltonian via  $b_{k\bar\alpha}^\dagger \to b_{k\bar\alpha}^\dagger \otimes d_{\bar\alpha}^\dagger$.
In all what follows, we consider the weak coupling regime and derive  a  Lindblad master equation in Born, Markov, 
and secular (BMS) approximation describing the time evolution of the system ($N$ two-level medium).
The system state $\rho^{(n)} \equiv \bra{n}\rho\ket{n}$ is conditioned upon the number of particles 
measured in the detector $n$.
The master equation assumes the form
$\dot{\rho}^{(n)}= \mathcal{L}_0 \rho^{(n)} + \mathcal{L}_+ \rho^{(n-1)} + \mathcal{L}_- \rho^{(n+1)}$, 
where jumps into (from) the bath $\bar\alpha$ are described by the super-operator $\mathcal{L}_+$ ($\mathcal{L}_-$)
and the probability of having counted $n$ particles after time $t$ is given by $P_n(t)= \trace{\rho^{(n)}(t)}$.
These $n$-resolved master equations are Fourier-transformed by introducing a counting field $\chi$ via $\rho(\chi) \equiv \sum_n \rho^{(n)} e^{in\chi}$, 
and similar for further detectors possibly placed in the other baths, leading to
a generalization of the master equation for collective spontaneous emission and re-absorption~\cite{agarwal},
 \bea\label{Eliouvillian}
 \dot{\rho} &=&{\cal L}(\{\chi_\alpha\}) \rho = -i \frac{\Omega}{2}\left[J^z,\rho\right]\nn
 &&+\sum_\alpha \Gamma_\alpha n_\alpha \left[e^{-i\chi_\alpha} J^+ \rho J^- - \frac{1}{2} \left\{J^- J^+, \rho\right\}\right]\nn
 &&+\sum_\alpha \Gamma_\alpha \left[1+n_\alpha\right]\left[e^{+i\chi_\alpha} J^- \rho J^+ - \frac{1}{2} \left\{J^+ J^-, \rho\right\}\right]\,,
 \eea
where we have omitted the dependence of $\rho$ on the counting fields for brevity and $\Gamma_\alpha\equiv 2\pi\sum_k \abs{h_{k\alpha}}^2 \delta(\Omega-\omega_{k\alpha})$ denotes the 
spontaneous photon emission rate of a single two-level system into reservoir $\alpha$, for which we
 assume stationary occupations $n_\alpha\equiv  \langle  b_{k\alpha}^{\dagger} b_{k\alpha} \rangle $ of photons with frequency 
 $\Omega=\omega_k$. 
 In the case of thermal photon reservoirs,  one has $n_\alpha=\left[e^{\beta_\alpha \Omega}-1\right]^{-1}$. 

 For a single counting field $\chi$,
 cumulants (denoted with double brackets throughout) of the photo-detection probability distribution $P_n(t)$ of 
 $n$ photons after time $t$ are obtained as
 $\dexpval{n^k(t)} \equiv \left(-\ii\partial_\chi\right)^k \mathcal{C}(\chi,t)|_{\chi=0}$,
 where $\mathcal{C}(\chi,t)\equiv \ln \traceS{{\rho}(\chi,t)}=\ln\traceS{e^{{\cal L}(\chi) t} \rho_0}$ 
 is the cumulant generating function.
%
%
\section{Results}

 \subsection{Single thermal reservoir}
 For a single ($N_\alpha=1$) 
 thermal bath at inverse temperature $\beta$, the Liouvillian $\mathcal{L}(\chi=0)$ 
  fulfills detailed balance and leads to equilibration of the system temperature with $\beta^{-1}$:
 the stationary state of Eq.(\ref{Eliouvillian}) 
 is simply given by
 $\bar \rho \propto e^{-\beta \Omega/2 J^z}$.
 In the angular momentum eigenbasis ($J^z \ket{jm} = 2 m \ket{jm}$ with $j=N/2$ and $-N/2\le m \le+N/2$), 
 $\mathcal{L}$

 has tridiagonal structure, and
 we obtain the FCS for transient Dicke super-radiance~\cite{GH82} 
 with all $N$ atoms initially in the excited state.
 At large times $t$,
 the FCS is given by the probability of $n$ photon emission and $N-n$ absorption events, i.e., 
 the distribution (for $0\le n \le N$) approaches
 $P_{n}(t\to \infty)\propto \left[1+n_B\right]^n n_B^{N-n}$ 
 with stationary moments  $\expval{\bar{n}^k} \equiv \lim\limits_{t\to\infty}\left(-\ii\partial_\chi\right)^k \traceS{{\rho}(\chi,t)}|_{\chi=0}$
 \bea
 \expval{\bar{n}^k}=\frac{\left[\sum_{m=1}^N m^k e^{m \beta \Omega}\right]}{\left[\sum_{m=0}^N e^{m \beta \Omega}\right]} 
 = \expval{\left[N-\left(J^z/2+N/2\right)\right]^k}_{\bar\rho},
 \eea
 which explicitly shows that the statistics of emitted photons and the statistics of 
 excitations remaining in the thermalized system are perfectly correlated.
 In a vacuum bath ($\beta\Omega\to\infty$), all $N$ initial excitations simply tunnel out of the system.
 For very large bath occupations ($\beta\Omega\to 0$) however, the corresponding higher cumulants grow super-linearly 
 with the system size 
 $\dexpval{\bar{n}}=N/2$, $\dexpval{\bar{n}^2}=N(N+2)/12$, $\dexpval{\bar{n}^4}=-N(N+2)\left[N(N+2)+2\right]/120$ and so on
 (all odd higher cumulants vanish).
 This super-linear scaling clearly is a collective effect: If the $N$ two-level atoms decayed independently, all cumulants
 would scale linearly in $N$ (cumulants of independent processes are additive).
Measuring these higher cumulants would require to detect all emitted photons with high accuracy.

 For measurements of transient FCS at finite times $t<\infty$, however, it is also necessary to consider 
 the finite bandwidth of the photo-detector given by its 
 time resolution $\Delta t$.
 The probability of detecting $n$ photons  
 during the time 
 interval $[t,t+\Delta t]$ is 
 then obtained via the inverse Fourier transform
 \mbox{$P_n^{\Delta t}(t) = \frac{1}{2\pi}\int\limits_{-\pi}^{+\pi}\trace{e^{{\cal{L}}(\chi)\Delta t - i n \chi} e^{{\cal L}(0) t} \rho_0} d\chi$}.
 We have used this expression to evaluate the fate of the Dicke super-radiance flash $\frac{d}{dt}\expval{n(t)}$ at finite time resolution $\Delta t$ (not shown).
 For $\Delta t \gtrsim \Gamma^{-1}$ (where $\Gamma$ is the single atom emission rate),
 the detector will essentially integrate over the radiation flash such that the signatures of collective emission can no longer be resolved in time. 
 In the following, we will argue that this situation drastically improves 
 in steady-state super-radiance (or more generally, stationary collective transport).
%
 %
 \subsection{Stationary transport and  `super-transmittance'} 
We now consider the case of stationary photon transport between $N_\alpha=2$ photon reservoirs $\alpha\in\{S,D\}$ (source and drain).  
 First of all, even when the reservoirs have non-thermal occupations $n_S\ge n_D$, the stationary  state of the medium
 (defined by ${\cal L}(0)\bar\rho=0$) 
 is a thermal state, $\bar\rho\propto e^{-\frac{\beta_M\Omega}{2}J^z}$, with  
 a 
 temperature $\beta_M^{-1}$  
 that corresponds to coupling to a  
 single fictitious ($N_\alpha=1$) reservoir 
 with occupation
 \bea\label{Eoccaverage}
 n_M=\frac{\Gamma_S n_S+\Gamma_D n_D}{\Gamma_S + \Gamma_D} \equiv \frac{1}{e^{\beta_M\Omega}-1}\,.
 \eea
 Such effective thermalization relations hold quite generally for effective tridiagonal rate equations when the system only
 supports a single transition frequency~\cite{schaller2011a}.
 Specifically, in the high-temperature limit $\beta_\alpha\Omega\ll 1$, this reduces to
 $T_M=(\Gamma_S T_S+\Gamma_D T_D)/(\Gamma_S+\Gamma_D)$, consistent with previous findings for coupled
 harmonic oscillators~\cite{Segal08}.
 
 Taking advantage of Eq.(\ref{Eoccaverage}), the stationary photon current $I_N$
 from the source bath through the coherent $N$-level medium into the drain bath may be evaluated conveniently
 in the angular momentum basis ($\rho_m\equiv\bra{jm}\rho\ket{jm}$), where one has $\bar{\rho}_{m+1}=\bar{\rho}_m n_M/(n_M+1)$.
 From the moment-generating function of the drain we obtain (note that trace conservation implies $\trace{{\cal L}(0) A}=0$ for arbitrary operators $A$)
 \bea\label{Ecurrent}
 I_N &\equiv & \lim\limits_{t\to\infty}\frac{d}{dt}\expval{n(t)} =\left.(-\ii \partial_{\chi})\trace{\mathcal{L}(\chi)e^{\mathcal{L}(\chi) t}\bar{\rho}}\right|_{\chi=0} =(-\ii) \traceS{{\cal L}'(0) \bar\rho}\nn
 &=& \Gamma_D(1+n_D) \traceS{J^-\bar\rho J^+} - \Gamma_D n_D \traceS{J^+\bar\rho J^-}\nn
 &=& \left(n_S-n_D\right) \frac{\Gamma_S \Gamma_D}{\Gamma_S + \Gamma_D} \sigma_N\,,
\eea
with
\bea
 \sigma_N&\equiv&
 \frac{\left(N - 2 n_M\right) \left(1 + n_M\right)^{N+1} 
 + n_M^{N+1}\left(2 + N + 2 n_M\right)}{\left(1+n_M\right)^{N+1} - n_M^{N+1}}\,.
 \eea
 This is one of our central results. First, for $N=1$ we obtain\\
\mbox{$\sigma_1 = \frac{\Gamma_S + \Gamma_D}{\Gamma_S(1+2 n_S)+\Gamma_D(1+2 n_D)}$}, coinciding with e.g. Ref.~\cite{Segal05},
 which in the linear response limit $\delta n \equiv n_S-n_D \to 0$ with a vacuum drain reservoir ($n_D=0$) reduces to unity. In this limit, we find
 the linear `photon-conductance' $\lim\limits_{\delta n\to0} \frac{I_1}{\delta n} = \gamma_{\rm cl} $, where 
 \bea 
 \gamma_{\rm cl}\equiv \frac{\Gamma_S \Gamma_D }{\Gamma_S+\Gamma_D}
 \eea
 is the classical transfer rate
 between reservoirs coupled in series via a medium with a single excitation.

 Second, for $N\ge 1$ and finite $n_S,n_D\ge 0$, the dimensionless quantity $\sigma_N$   describes the {\em degree of collectivity}
 of the steady-state transport process between $S$ and $D$. 
 In the large $N$-limit, $\sigma_N$ is asymptotically given by a single parameter scaling form,
 \bea
 \lim_{N\to \infty}\frac{\sigma_N}{N} &=& \coth\left[\frac{1}{2} \frac{N}{n_M}\right] - 2 \frac{n_M}{N}\,,
 \eea
 from where the main limits of the current are easily accessible:
 when the 
 occupation parameter $n_M$ (Eq.(\ref{Eoccaverage})) is 
 much
 smaller than the effective system size $N$ (at constant density of two-level systems in the medium),  i.e. for $n_M/N \ll 1$,
 we have $\sigma\approx N$.
 In this limit, the current $I_N$ is an incoherent sum of independent contributions from $N$ classical transfer processes with rates $\gamma_{\rm cl}$ 
 and thus scales linearly with $N$.

 In contrast, for $n_M/N \gg 1$ the reservoirs provide sufficiently large photon flux to drive the medium 
 into the regime of collective, steady-state photon `super-transmittance'.
 There, one obtains to leading order $\sigma_N\approx \frac{N^2}{6 n_M}$, such that the current reads
 \bea
 I_N \approx \gamma_{\rm cl} \frac{n_S-n_D}{n_M} \frac{N^2}{6}
 \eea
 with the quadratic dependence in $N$ indicating the collective nature of the transmission process. 
 Note that when $n_D\to0$ (and $\Gamma_S>0$), this further reduces to $I_N\approx \Gamma_D N^2/2$, which gives the 
 theoretical 
 maximum of the 
 current in the collective transport regime (the independence of the source tunneling rate $\Gamma_S$ is a consequence of the required 
 limit of $n_S\to\infty$ here). 
 Super-transmittance also occurs for transport between two {\em thermal} reservoirs at sufficiently large temperatures  $T_D=T$ and $T_S=T+\Delta T \gg \hbar\Omega/k_B$.
 For $\Delta T \to 0$, the classical thermal conductance $k_{\rm th}\equiv \lim_{\Delta T \to 0} \Omega I_N/ \Delta T $ through the medium 
 is then obtained  from Eq.(\ref{Ecurrent}) and again scales quadratically with $N$, 
 $k_{\rm th} = \gamma_{\rm cl}  \left(\frac{\Omega}{T}\right) N(N+2)/6$, whereas at small temperatures $T\ll \hbar\Omega/k_B$ one has
 $k_{\rm th} = \gamma_{\rm cl}  \left(\frac{\Omega}{T}\right)^2 N e^{- \Omega/T} $. 

 For $N\ge 4$, the long-term limit of the cumulant-generating function for the current
 (given by the dominant eigenvalue of ${\cal L}(\chi)$ as $\chi\to 0$) cannot be determined analytically anymore.
 To analyze its global analytic features we therefore consider the characteristic polynomial 
 $\abs{{\cal L}(\chi)-\lambda \f{1}}\equiv f(\chi)$, for which we find the symmetry
 $f(\chi)=f\left(-\chi-\ii \ln\left(\frac{n_D(1+n_S)}{n_S(1+n_D)}\right)\right)$.
 This analytic symmetry transfers to the dominant eigenvalue,
 such that we may deduce for thermal baths the steady state fluctuation theorem~\cite{esposito}
\bea
\lim_{t \to \infty} \frac{P_n(t)}{P_{-n}(t)}=e^{\Omega (\beta_S - \beta_D) n}
\eea
for arbitrary system size $N$, linking the probabilities of $n$ tunneled particles after time $t$ with both reservoir
temperatures.
%
 \subsection{Higher photon current cumulants}
 We now discuss our second key result
 concerning the non-trivial scaling behavior with $N$ of the higher stationary photon current cumulants
$ C_{k}\equiv \lim\limits_{t\to \infty} \frac{d}{dt} \dexpval{n^k(t)}$
 at $k>1$ 
 in regimes where the stationary photon current $I_N=C_1$ itself is less suitable to indicate collective features.
 
 For large times, particle cumulants $\dexpval{n^k(t)}$ grow linearly, such that they are related to the current cumulants via
 $\dexpval{n^k(t)} \to C_k t + S_k$ with a constant shift term $S_k$. 
 Already for moderate system sizes the cumulant-generating function cannot be determined directly anymore, calculating the Laplace
 transform of the moment-generating function $\tilde{\cal M}(\chi,z) = \trace{\left[z\cdot\f{1}-{\cal L}(\chi)\right]^{-1}\bar{\rho}}$ 
 however only requires the inversion of an $(N+1)\times(N+1)$ matrix.
 Therefore, we make use of the relation between cumulants and moments: As the $n$-th moment is given by a polynomial of order $n$
 in the first $n$ cumulants, we obtain e.g. for the third cumulant
 \mbox{$\dexpval{n^3(t)}=\expval{n^3(t)}-3 \dexpval{n^1(t)} \dexpval{n^2(t)} - \dexpval{n^1(t)}^3$}, 
 which in the long-term limit reduces to
 \mbox{$C_3 \to \lim\limits_{t\to\infty}\frac{d}{dt}\left[\expval{n^3(t)}-3 \left(C_1 t + S_1\right) \left(C_2 t + S_2\right) - \left(C_1 t + S_1\right)^3\right]$}.
 Using limit properties of the Laplace transform, we may therefore directly extract the current cumulants $C_k$ once all smaller cumulants and the Laplace
 transform of the $k$-th  moment is known.

 The most obvious regime is $\delta n = 0$, where there is no net stationary transport
 through the system. 
 In fact, we find that all odd cumulants $C_{{2k+1}}$ vanish.
 Still, the two reservoirs (with environmental photon occupations $n_S=n_D \equiv n_E$ at the systems transition frequency $\Omega$)
 induce stationary fluctuations (thermal fluctuations for thermal occupation numbers $n_E$).
 Explicitly, for small $n_E$ and symmetric coupling $\Gamma_S=\Gamma_D\equiv \Gamma$ for simplicity we find 
 \bea
 C_2 &=& \frac{\Gamma  n_E(1+n_E)}{(1+n_E)^{N+1}-n_E^{N+1}} \left[N + (N+2)(N-1) n_E + \ord{\{n_E^2\}}\right]\,,\nn
 C_4 &=& \frac{\Gamma  n_E(1+n_E)}{\left[(1+n_E)^{N+1}-n_E^{N+1}\right]^3}\left[N + \left(3N(N+2)-8\right) n_E + \ord{\{n_E^2\}}\right]\,,\nn
 \eea
 where the quadratic scaling corrections $\propto N^2$ underline the collective effect of the fluctuations.
 In the opposite case, where $n_E$ is large, the fluctuations behave asymptotically as
 \bea
 C_2 &\to& \frac{\Gamma  n_E}{6} N (N+2)\,,\nn
 C_4 &\to& \frac{\Gamma  n_E}{360} N (N+2) \left[N (N+2) + 12\right]\,,
 \eea
 i.e., the super-linear scaling of higher cumulants in the transient Dicke effect transfers to the stationary regime.
 In essence, this already demonstrates that even in regimes, where the current does not reveal any collective 
 signatures, higher cumulants may do so.

\begin{figure}[t]
\begin{center}
 \includegraphics[width=0.8\textwidth,clip=true]{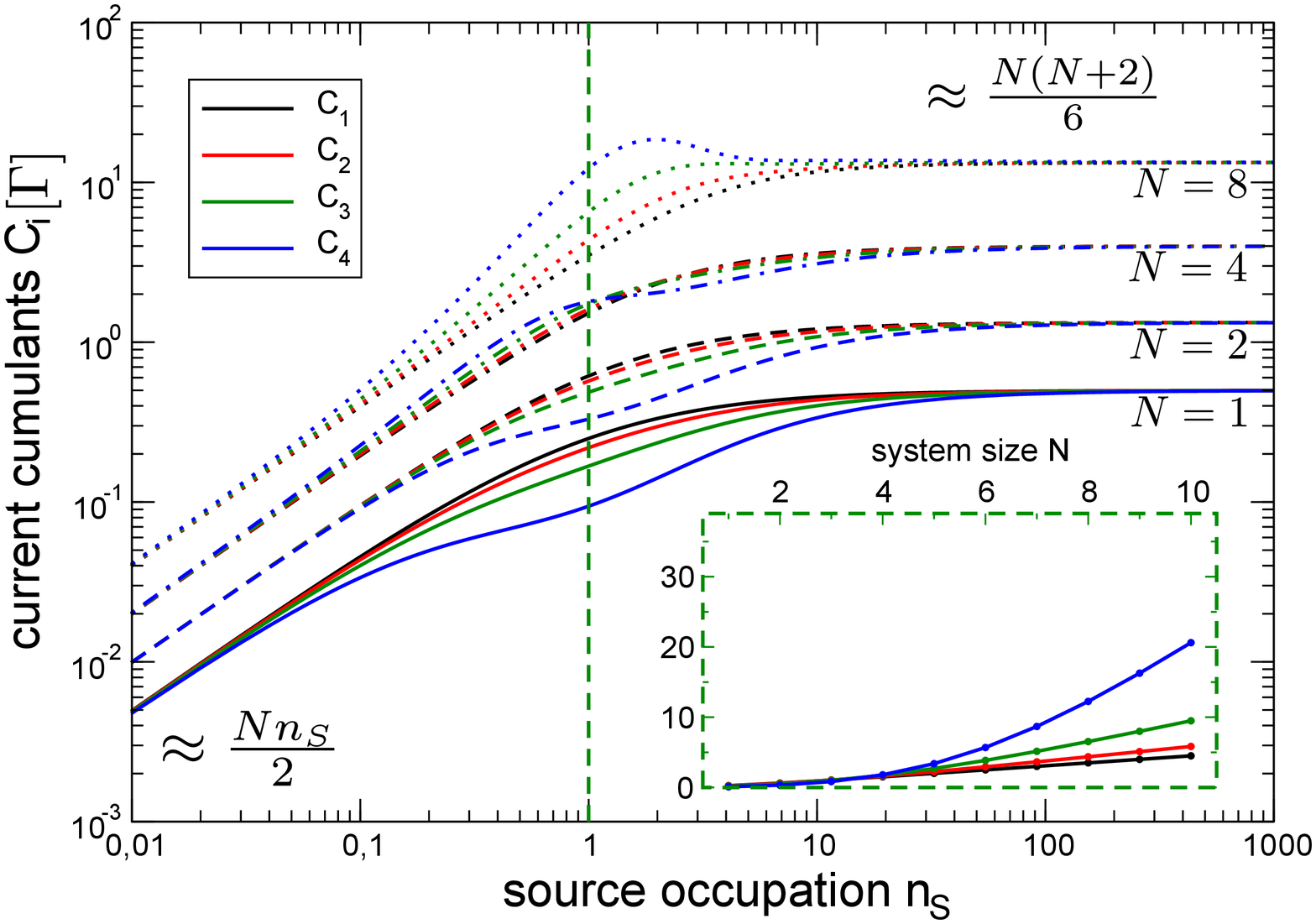}
\end{center}
 \caption{\label{Fcumulants}(Color Online)
 Logarithmic plot of the first few current cumulants $C_1$ (black), $C_2$ (red), $C_3$ (green), and $C_4$ (blue) in units of $\Gamma=\Gamma_S=\Gamma_D$ versus the source occupation $n_S$ for 
 different system sizes $N=1$ (solid), $N=2$ (dashed), $N=4$ (dash-dotted) and $N=8$ (dotted).
 The first cumulant shows a relatively smooth transition from linear (small differences for small $n_S$) to quadratic (large differences for large $n_S$) scaling 
 with the system size $N$.
 The higher cumulants however show super-linear scaling behavior in a regime where the current does not (compare the inset taken at the dashed line $n_S=1$).
 Other parameters: $\Gamma_S=\Gamma_D=\Gamma$, $n_D=0$.
 }
 \end{figure}

 We have also derived analytical expressions for the cumulants in the cases of non-vanishing but small and very large bias
 $\delta n \equiv n_S-n_D$,  with numerical results for finite $\delta n$ and $n_D=0$ displayed in Figure~\ref{Fcumulants}. 
 For vanishing drain occupation $n_D=0$ and small bias $\delta n=n_S$, the cumulant-generating function scales as 
 ${\cal C}(\chi,t) = \gamma_{\rm cl} (e^{i\chi}-1) N n_S t$, which yields a Poissonian process with cumulants $C_k=\gamma_{\rm cl} N n_S$.
 In contrast, in the large bias limit  $\delta n\to \infty$, the FCS  
 depends quadratically on $N$ also for finite $n_D$, and we recover the collective transport regime:
 The corresponding cumulant-generating function then reads 
\bea
{\cal C}(\chi,t)&=&\frac{\Gamma_D}{6} N (N+2) \left[\left(e^{+i\chi}-1\right)\left(1+n_D\right) + \left(e^{-i\chi}-1\right) n_D\right] t\,,
\eea
which yields the photon current cumulants
$C_{k} =  \Gamma_D  N(N+2)/6 \left[1+2 n_D \delta_{k, {\rm even}}\right]$
and describes the independent superposition of two counter-propagating Poissonian processes (with identical cumulants) corresponding to absorption and emission.
 One recognizes in Figure~\ref{Fcumulants} that indeed higher cumulants yield an earlier onset of super-linear scaling behavior with the system size $N$, as compared with
 the stationary photon current $I_N=C_1$. 
 \section{Experimental verification}
 Finally, we propose an experimental setup for testing our predictions:
 In order to achieve the small sample limit of super-radiance, it is necessary to constrain the size of the optically active medium to
 dimensions smaller than the photon wavelength. We therefore propose a gas of $N$ atoms confined in a cylindrically shaped nano-cell~\cite{Sarkisyanetal04}, where $N$ is 
 controlled, e.g.,  by evaporation from a particle reservoir. We recall that the cross-over between collective $(I_N\propto N^2)$ and non-collective  $(I_N\propto N)$ photon current regimes is 
 (within the limits of our model Hamiltonian)
 solely controlled by a single parameter,  $n_M$ in Eq.(\ref{Eoccaverage}). For thermal sources, to achieve the collective regime where $n_M\gg N$ would
 require large source occupations $n_S(\Omega)$ and thus very large temperatures $k_B T \gg \hbar \Omega$.
Therefore, in order to reach this regime we suggest as source an optical cavity with loss rate $\kappa$ and pumped by a laser at frequency $\omega_{\rm L}$ with pump strength $p$, compare  Figure~\ref{fig:sketch}  (b).  
The corresponding stationary cavity photon distribution is $n_S(\Omega) = \frac{\abs{p}^2}{\left(\Omega-\omega_{\rm L}\right)^2+\kappa^2}$.
Changing e.g. the detuning of the laser then enables one to change $n_S(\Omega)$ and thereby to switch between the different scaling 
 regimes discussed above. The nano-cell is placed within the cavity, and in directions perpendicular to the cell axis, the atoms  are then collectively driven by the cavity modes and collectively 
 emit into (vacuum) non-cavity modes which have $n_D(\Omega)\approx 0$ at optical transition frequencies. 
The photon emissions are counted and time-resolved by a detector whose signal is then used to construct the distribution $P_n(t)$. Sampling a longer signal over different initial times then allows one to efficiently construct also higher cumulants $C_k$~\cite{Flietal09}.

\section{Summary}
To conclude, we have demonstrated that the transient Dicke super-radiance may be transferred to a stationary regime. There, 
it is reflected as a super-linear scaling of transport cumulants with the system size, which we have termed super-transmittance.
Both in transient and stationary setups, higher cumulants yield a much more sensitive signature of collective effects than the first cumulant.
The effects should be measurable with present techniques.
%
\section*{Acknowledgements}
The authors gratefully acknowledge support by the DFG in the
research training group GRK 1558 (M. V.), grants SCHA 1646/2-1 (G. S.), BRA 1528/7, BRA 1528/8, SFB 910 (T. B.). We 
have significantly benefited from valuable suggestions by F. Renzoni and  discussions with V. Bastidas, F. Brennecke, M. Esposito, and G. Kie{\ss}lich.
%


%
%
\end{document}